\documentstyle[12pt]{article}
\newcommand{\dirak}{ /\!\!\!\!{\cal D}}
\newcommand{\be}{\begin{equation}}
\newcommand{\ee}{\end{equation}}
\newcommand{\ba}{\begin{eqnarray}}
\newcommand{\ea}{\end{eqnarray}}
\newcommand{\Lgg}[1]{\ln^{#1}\frac{\Lambda^2}{M_0^2}}
\newcommand{\lgg}[1]{\ln^{#1}\frac{\Lambda^2}{\mu^2}}
\newcommand{\acc}[1]{\left[
1+ O\left(\frac{1}{\ln^{#1}\frac{\Lambda^2}{\mu^2}}\right)\right]}
\newcommand{\nl}{\nonumber \\}
\begin{document}
\begin{center}
{\Large\bf
Composite two-Higgs model with dynamical CP-violation.}\\
{\bf A.A.Andrianov, V.A.Andrianov, V.L.Yudichev}\\
Sankt-Petersburg State University

\end{center}

\begin{abstract}
Quark models with four-fermion interaction including derivatives of fields
are exploited as prototypes for composite-Higgs extensions of the Standard
Model. In the non-trivial case of two- and four-derivative insertions the
dynamical breaking of chiral symmetry occurs in two channels, giving rise
to two composite Higgs doublets. For special configuration of four-fermion
coupling constants  the dynamical CP-violation in the Higgs sector appears
as a result of complexity  of two  v.e.v. for Higgs doublets. In this
scenario the second Higgs doublet is regarded as a radial excitation of
the first one.
\end{abstract}

\section{Introduction.}

Higgs particles have been introduced in  the Standard Model
of electroweak interaction (SM) in
order to get massive vector bosons as the mediators of
electroweak interaction without loss of renormalizability
or unitarity of the
theory. They are indispensable in the modern picture
of the particle world \cite{Gunion},\cite{Bardeen},
though being not yet detected experimentally.

In certain approaches the quark self-interaction
is responsible for  production of quark-antiquark
bound states which are identified as
composite Higgs particles.
In a minimal version, quark models with local four-fermion interaction
 are used to
derive an effective action for  Higgs doublets
(Top-condensate SM \cite{Bardeen}).
Following this way, we propose the quark model
with quasilocal four-fermion interaction
\cite{4ferm}where the
derivatives of fermion fields are included to allow the formation of
the second Higgs doublet. We consider two Higgs doublets in the model
\cite{Lee}
because one doublet does not provide any effect of
dynamical CP-violation. We are going to show how P-parity
breaks down dynamically
for the special choice of coupling
constants \cite{CP} of the four-fermion interaction.

\section{Quark model with quasilocal interaction.}

For simplicity, we restrict ourselves with a two-flavor
quark model in which the
$ t$- and $ b$-quarks are involved in the
Spontaneous Breaking of Chiral Symmetry (SBCS).
In accordance with SM, the  left components of both  quarks form
a doublet:
\be
  q_{L}=\left( t_{L} \atop b_{L}\right)
\ee
which transforms under
$ SU(2)_{L} $ group as a fundamental representation
while the right components
$ t_{R}, b_{R} $ are singlets.
The lagrangian for a minimal model which leads to the strong coupling
regime is taken in the form:

\ba
{\cal L_{J}}&=&\bar q_{L}\dirak q_{L} + \bar t_{R}\dirak t_{R}+
  \bar b_{R}\dirak b_{R}+\nl&&
\frac{8\pi^2}{N_{c}\Lambda^{2}}\sum_{k,l=1}^{2}a_{kl}\left(
g_{t, k}J^{T}_{t, k}-g_{b, k}\widetilde J^{T}_{b, k}\right) i\tau_{2}
\left(g_{b, l}J_{b, l}-g_{t, l}\widetilde J_{t, l} \right). \label{lag}
\ea
Here we introduced the following
denotations for doublets of fermion currents:
\be
  J_{t, k}\equiv \bar t_{R}\varphi_{k}\left(
-\frac{\partial^{2}}{\Lambda^{2}} \right)q_{L} , \qquad
  J_{b, k}\equiv \bar b_{R}\varphi_{k}
\left(-\frac{\partial^{2}}{\Lambda^{2}} \right)q_{L},\label{J}
\ee
The tilde in
$\widetilde J_{t, k}$  and  $ \widetilde J_{b, k} $ marks
charge conjugated quark currents:
\be
\widetilde J_{t, k}=i\tau_{2}J^{\star}_{t, k}, \qquad
\widetilde J_{b, k}=i\tau_{2}J^{\star}_{b, k}
\ee
 The subscripts
$ t, b$ indicate right components of
$ t $ and $ b $ quarks in the currents, the index
$ k$ enumerates the formfactors:
\be
  \varphi_{1}=1,\quad \varphi_{2}=\sqrt{3}\left(1-2
\left(-\frac{\partial^{2}}{\Lambda^{2}} \right)\right).
\ee
 As the spinor
indices are contracted to each other in  (\ref{J}),
$ J_{t, k} $ transforms as a doublet under
$ SU(2)_{L} $.
$ \tau_{2} $ is a Pauli matrix in the adjoint representation
of the group
$ SU(2)_{L} $.
Coupling constants of the four-fermion interaction are represented by
$ 2\times 2 $ matrix
$ a_{kl} $.  We introduce also
the Yukawa constants $ g_{k, l}, g_{b, k} $.

Insofar as the theory with four-fermion  interaction is nonrenormalizable,
it is understood as an effective low-energy field theory
with  a cut-off $ \Lambda $
at large momenta
in Euclidean metric.

\section{Higgs sector.}

The lagrangian
to describe the dynamics of composite Higgs bosons
can be obtained by means of
introduction of auxiliary bosonic variables and by integrating out
fermionic degrees of freedom. According to this programme, we
define two doublets:
\be
  H_{1}=\left(h_{11}\atop h_{12}\right),\qquad
  H_{2}=\left(h_{21}\atop h_{22}\right)
\ee
and their charge conjugates:
\be
  \widetilde H_{1}=\left(h_{12}^{\star}\atop -h_{11}^{\star}\right),\qquad
  \widetilde H_{2}=\left(h_{22}^{\star}\atop -h_{21}^{\star}\right).
\ee
In terms of auxiliary fields, the lagrangian (\ref{lag}) can be
rewritten in the following way:
\be
\frac{N_{c}\Lambda^{2}}{8\pi^2}\sum_{k,l=1}^{2}H^{\dagger}_{k}(a^{-1})_{kl}H_{l}+
\sum_{k=1}^{2}\left[
g_{t, k}\widetilde H_{k}^{\dagger}J_{t, k}+
g_{b, k}H_{k}^{\dagger}J_{b, k}\right] + h.c.
\ee
The integrating out of  fermionic degrees of freedom will produce
the effective action for Higgs bosons of which we shall keep  only the
kinetic term and  the effective potential
consisting of two- and four-particles
vertices.
The omitted terms are supposedly small,
being proportional to inverse powers of a large scale factor.

The effective potential which does not violate the CP-parity
manifestly is
parametrized by seven coupling constants
$ f_{m} $
\ba
V_{eff}=\frac{N_{c}}{8\pi^2}\Biggl[
-\sum_{k,l=1}^{2}\Delta_{kl}\left(H^{\dagger}_{k}H_{l}\right)+
f_{1}\left(H_{1}^{\dagger}H_{1}\right)^{2}+\nl
f_{2}\left(H_{1}^{\dagger}H_{1}\right)\left(H_{1}^{\dagger}H_{2}\right)+
f_{3}\left(H_{1}^{\dagger}H_{1}\right)\left(H_{2}^{\dagger}H_{2}\right)+\nl
f_{4}\left(H_{2}^{\dagger}H_{2}\right)\left(H_{1}^{\dagger}H_{2}\right)+
f_{5}\left(H_{2}^{\dagger}H_{2}\right)^{2}+\nl
f_{6}\left(H_{1}^{\dagger}H_{2}\right)^{2}+
f_{7}\left(H_{1}^{\dagger}H_{2}\right)\left(H_{2}^{\dagger}H_{1}\right) + h.c.
\Biggr],\label{pot}
\ea
where the mass term is in general nondiagonal  and  represented
by
$ 2\times 2 $ matrix
$ \Delta_{kl} $.

We assume the vacuum charge stability or, in other words,
that only  neutral components  of both Higgs doublets
may have nonzero v.e.v. Hence, one
can deal with only neutral components of the Higgs doublets
in the effective action for studying SBCS.
 This part of the Higgs sector
 can be investigated separately as a model where two singlets
(not doublets) appear as composite Higgs bosons.
For this purpose, we  use the two-channel
 model which we have already developed for the case of one-flavor.
\cite{2channel}

Further on, we restrict ourselves with two choices of Yukawa constants
$ g_{t, l}, g_{b, l} $: 1)
$ g_{t, k}=g_{b, k} $ and 2)
$ g_{t, k}\gg g_{b, k} $.
Let us proceed to demonstration of the dynamical P-parity
breaking in the one-flavor quark model.

\section{Dynamical breaking of P-parity.}

The one-flavor quark model with quasilocal four-fermion
interaction is considered near its polycritical point
\cite{CP},\cite{2channel}:
\be
      a_{ij}\sim\delta_{ij}+\frac{\Delta_{ij}}{\Lambda^2}, \qquad
      |\Delta_{ij}|\ll \Lambda^2,                            \label{11}
      \ee
where the elements of the matrix $ \Delta $ parametrize
deviation from a polycritical point  in the
three-dimensional coupling constants space.
Following the definitions made in \cite{CP},  we relate the fields
$ \chi_{1} $ and
$ \chi_{2} $ to the neutral components of Higgs doublets.
The potential (\ref{pot}) in this case  transforms into:
\ba
V_{eff}&=&\frac{N_{c}}{8\pi^2}\Biggl[
-\sum_{k,l=1}^{2}\Delta_{kl}\left(\chi_{k}\right)+
f_{1}|\chi_{1}|^{4}+
f_{2}|\chi_{1}|^{2}\chi_{1}^{\star}\chi_{2}+\nl &&
\hat f_{3}|\chi_{1}|^{2}|\chi_{2}|^{2}+
f_{4}|\chi_{2}|^{2}\chi_{1}^{\star}\chi_{2}+
f_{5}|\chi_{2}|^{4}+
f_{6}(\chi_{1}^{\star}\chi_{2})^{2}+h.c.
\Biggr],
\ea
where
$ \hat f_{3}=f_{3}+f_{7} $, and
 one obtains from the model:\\
for the case 1)
\be
f_{1}=4\log\frac{\Lambda^{2}}{\mu^2},\quad
f_{2}=\frac{3\sqrt{3}}{8},\quad
\hat f_{3}=\frac{3}{4},\quad
f_{4}=\frac{\sqrt{3}}{4},\quad
f_{5}=\frac{9}{16},\quad
f_{6}=\frac{3}{16},\quad
\ee
and for the case 2):
\be
f_{1}=8\log\frac{\Lambda^{2}}{\mu^2},\quad
f_{2}=\frac{3\sqrt{3}}{4},\quad
\hat f_{3}=\frac{3}{2},\quad
f_{4}=\frac{\sqrt{3}}{2},\quad
f_{5}=\frac{9}{8},\quad
f_{6}=\frac{3}{8}.\quad
\ee

As  complex solutions of the mass-gap equations are expected,
we specify for them an appropriate parametrization
which brings a real dynamical mass
$M_{0}=\chi_{1}+\sqrt{3}\chi_{2}$:
\be
\bar\chi_1=\chi_1+i\rho, \quad
\bar\chi_2=\chi_2-i\frac{\rho}{\sqrt{3}},\quad \chi_i\equiv\mbox{ Re}
\bar\chi_i,
\label{13}
\ee
where
$ \chi_{1}, \chi_{2} $ and
$ \rho $  are real.
In these variables the mass-gap equations read:
\ba
\Delta_{11}\chi_1+\Delta_{12}\chi_2&=&M_0^3\Lgg{} -6\sqrt{3}\chi_1^2\chi_2-
18\chi_1\chi_2^2-8\sqrt{3}\chi_2^3\nl
d_1\chi_1-d_2\chi_2&=&2\sqrt{3}\chi_1(\chi_1^2+3\chi^2_2)+2\rho^2(
\frac{4}{\sqrt{3}}\chi_1-2\chi_2)\nl
\rho(\sqrt{3}\Delta_{11}-\Delta_{12})&=&2\rho\sqrt{3}(\chi_1^2+\chi_2^2+
\frac43\rho^2), \label{14}
\ea
where
\be
d_1=\sqrt{3}\Delta_{11}-\Delta_{12},\quad\quad
d_2=-\sqrt{3}\Delta_{21}+\Delta_{22} \label{15}.
\ee
When the coupling constants of the quark model are chosen
near a particular hyperplane in the coupling constant
space,  defined by the equation
$ \sqrt{3}d_{1}-2d_{2}=0 $, the non-zero solution for
$ \rho $ exists, and in the large-log approximation
one has:
\be
\chi_1^2=\frac{d_1+4\Delta_{12}}{16\sqrt{3}(\lgg{}-3)} \qquad
\chi_2\approx -\sqrt{3}\chi_1,  \label{31}
\ee
\be
\rho^2=\frac{d_1\sqrt{3}}{8}-\frac34( \chi^2_1+\chi^2_2 )=
\frac{d_1\sqrt{3}}{8}
\acc{}.                             \label{31'}
\ee
\section{Second Variation and Mass Spectrum of Composite
States.}

The matrix of  second
variations of the effective potential
determines the spectrum of bosonic states. We divide it into two
parts: the first one independent on momentum, $\hat B$, and the kinetic part
 $\hat A p^2$.
\be
\delta^{(2)}S=(\delta\chi^{\star}, (\hat Ap^2+\hat B)\delta\chi)
\ee
\be
\chi_j=<\!\chi_j\!>+\delta\chi_j =<\!\chi_j\!>+
\sigma_j+i\pi_j.
\ee
The constant matrix $\hat B$ has the  zero-mode
$ \chi^0_j=<\!\pi_j\!>\!- i\cdot
<\!\sigma_j\!> $, regarding to the existence of Goldstone
bosons.
In order to find the spectrum of collective excitations one should solve
the equation
\be
\label{41} \det( \hat Ap^2+\hat B )=0.
\ee
 at $-m^2=p^2\le 0$.

One can see from (\ref{31}),(\ref{31'}) that in
the large-log limit the axial dynamical mass
 dominates. It leads to
appearance of a massless boson in the scalar channel
in accordance to the Goldstone theorem.

The classification of states given by P-parity quantum
number is relevant only in the large-log limit,
when:
\be
\frac{B^{\pi\sigma}}{B^{\sigma\sigma}}\approx
\frac{B^{\pi\sigma}}{B^{\pi\pi}}=O\left( \frac{1}{\lgg{}}
\right), \label{58}
\ee
next-to-leading logarithmic
effects are of no importance and one can neglect mixing of the
states with different P-parity.
Then the
spectrum of mesons is:
\begin{eqnarray}
  m_{\sigma}^2=0,&& m_{\sigma'}^2\approx
\frac{d_1+4\Delta_{12}}{\sqrt{3}\lgg{}}
\approx 16\chi_1^2=4M_c^2
\label{59}      \nl
m_{\pi'}^2\approx \sqrt{3}d_1,&& m_{\pi}^2\approx
\frac{4( d_1+\Delta_{12} )}{9\sqrt{3}\lgg{}}
\end{eqnarray}

The ratio of $m_{\sigma'}$ and $m_{\pi}$
does not depend on logarithm, therefore the masses are comparable.
On the other hand, in the
models with finite momentum cut-off,
the effects of order of
$ \Lambda^2 $ become important and
 the dynamical P-parity
breaking is induced, since $ B_{\pi\sigma}\not=0$.

Thus, we have constructed the model where:\bigskip \\
 a). Two Higgs doublets are created dynamically as a consequence
of SBCS in two channels.\bigskip \\
 b). The appopriate fine tuning leads also to spontaneous breaking of
P-parity and, therefore, of CP-parity in the Higgs sector.

 The experimental implications of such effects are expected
to be rather small  in the fermion sector of SM \cite{Gunion},\cite{Skjold}.
These effects are
observable only in decays of heavy Higgs particles (namely, pseudoscalar
Higgses may decay into scalar ones,  scalar Higgses may decay
into pseudoscalar ones) and in decays of Higgs particles into
two bosons where CP-even and CP-odd amplitudes appear.\bigskip

This paper is supported by the RFFI grant No. 95-02-05346-a and by
the GRACENAS grant No. 95-6.3-13

\end{document}